\documentclass[10pt,conference]{IEEEtran}
\ifCLASSINFOpdf
  \usepackage[pdftex]{graphicx}
  \usepackage[bookmarks=false]{hyperref}
  \usepackage{caption}
  \usepackage{subcaption}
  \usepackage{color}
\else
  \usepackage[dvips]{graphicx}
\fi
\usepackage[usenames,dvipsnames,svgnames]{xcolor}

\newcommand{\optional}[1]{\ignorespaces}

\begin{document}
%
\title{D-STREAMON: from middlebox to distributed NFV framework for network monitoring*}

\author{\IEEEauthorblockN{
Pier Luigi Ventre\IEEEauthorrefmark{1}\quad
Alberto Caponi\IEEEauthorrefmark{2}\quad
Giuseppe Siracusano\IEEEauthorrefmark{1}\quad
Davide Palmisano\IEEEauthorrefmark{1}\\
Stefano Salsano\IEEEauthorrefmark{1}\IEEEauthorrefmark{2}\quad
Marco Bonola\IEEEauthorrefmark{2}\quad
Giuseppe Bianchi\IEEEauthorrefmark{1}\IEEEauthorrefmark{2}
}
\IEEEauthorblockA{
\IEEEauthorrefmark{1}University of Rome Tor Vergata, Italy\\
\IEEEauthorrefmark{2}Consorzio Nazionale Interuniversitario per le Telecomunicazioni (CNIT), Italy}
Email: \{name.surname\}@uniroma2.it
}


%


\maketitle

\begin{abstract}
Many reasons make NFV an attractive paradigm for IT security: lowers costs, agile operations and better isolation as well as fast security updates, improved incident responses and better level of automation. On the other side, the network threats tend to be increasingly complex and distributed, implying huge traffic scale to be monitored and increasingly strict mitigation delay requirements. Considering the current trend of the networking and the requirements to counteract to the evolution of cyber-threats, it is expected that also network monitoring will move towards NFV based solutions. In this paper, we present D-StreaMon an NFV-capable distributed framework for network monitoring realized to face the above described challenges. It relies on the StreaMon platform, a solution for network monitoring originally designed for traditional middleboxes. An evolution path which migrates StreaMon from middleboxes to Virtual Network Functions (VNFs) has been realized. 
\end{abstract}

%
\IEEEpeerreviewmaketitle

\let\svthefootnote\thefootnote
\let\thefootnote\relax\footnotetext{*This research was partially supported by the EU Commission within the Horizon 2020 program, SCISSOR project \cite{scissor} grant no 644425.}
\let\thefootnote\svthefootnote

\vspace{1.0ex}

\vspace{-1.5ex}

\section{Introduction}
\IEEEpubidadjcol
The unyielding increasing trend, the evolving complexity and the diversified nature of modern cyber-threats \cite{isrt_symantec_2016} calls for new scalable, accurate and flexible solutions for network traffic monitoring and threats detection. New solutions should support zero-touch configurations, flexibility, agility and real-time traffic analysis capabilities to promptly detect and mitigate cyber-attacks. The real challenge is to promptly react to the high mutable needs by deploying customizable traffic analyses functionalities, capable of tracking events and detect different behaviours of attacks. Such objectives can be achieved by efficiently handling of the many heterogeneous features, events, and conditions which characterize an operational failure, a network's application mis-behavior, an anomaly or an incoming attack. The flexibility in the network monitoring can be obtained through the systematic exploitation of stream-based analysis techniques, carefully combined to address \textit{scalability-by-design} and made available to the programmer by means of open APIs. Even more challenging, traffic analyses and mitigation primitives should be ideally brought inside the monitored network and the monitoring probes themselves. This avoids the centralization of the analysis which moves flows of data to a central point resulting inadequate to cope with the huge traffic scale and the strict (ideally real-time) mitigation delay requirements. Solutions for the aforementioned challenges find breeding ground in the current trends towards the softwarization of networks~\cite{galis}~\cite{kind}. Such trends includes technologies like Network Function Virtualization (NFV)~\cite{nfv}, Software Defined Networking (SDN)~\cite{sdn} and Cloud Computing~\cite{cloud}. More specifically NFV, focusing on the decoupling of network functions from physical devices on which they run, becomes the ideal technology to bring to the reality the idea of pervasive network monitoring. NFV leverages on virtualization technologies to execute network appliances in virtual resources which are deployed in commodity servers. We expect that fields like network security could benefit in moving towards NFV paradigm, where security systems or hardware based middleboxes are substituted by virtualized network running in commodity servers.

\begin{figure}
\centering
\begin{subfigure}{0.24\textwidth}
  \centering
  \includegraphics[width=1.0\textwidth]{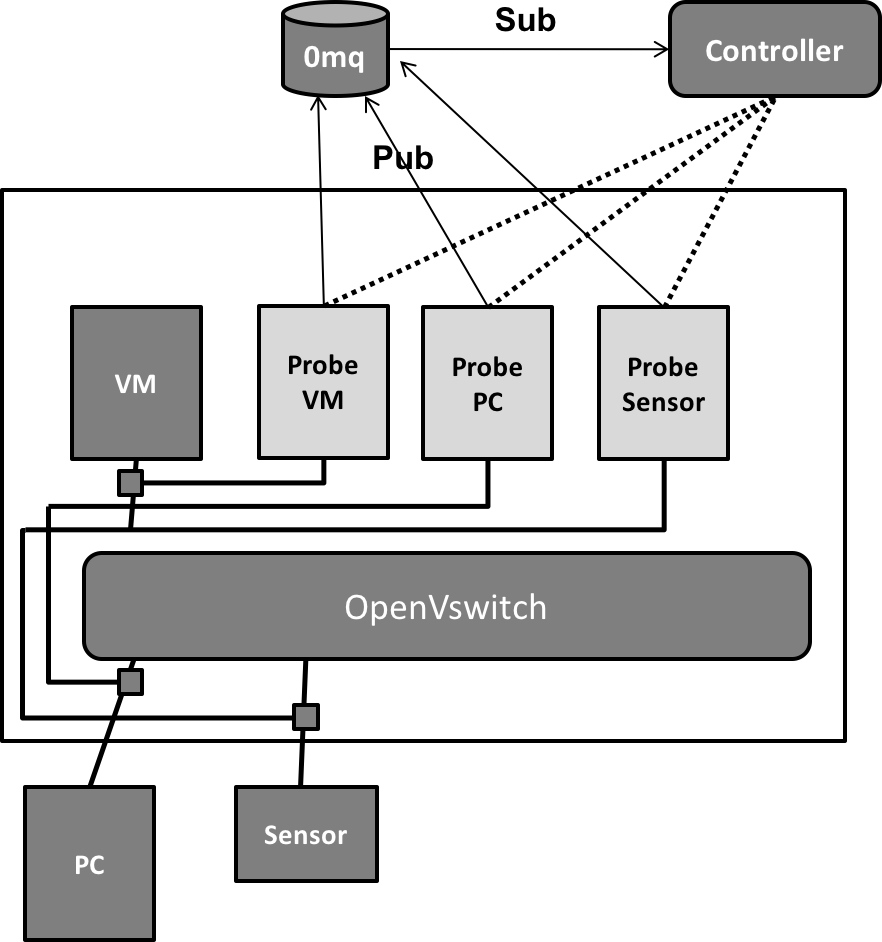}
  \caption{Probes as Processes}
  \label{fig:sub2}
\end{subfigure}
\begin{subfigure}{0.24\textwidth}
  \centering
  \includegraphics[width=1.0\textwidth]{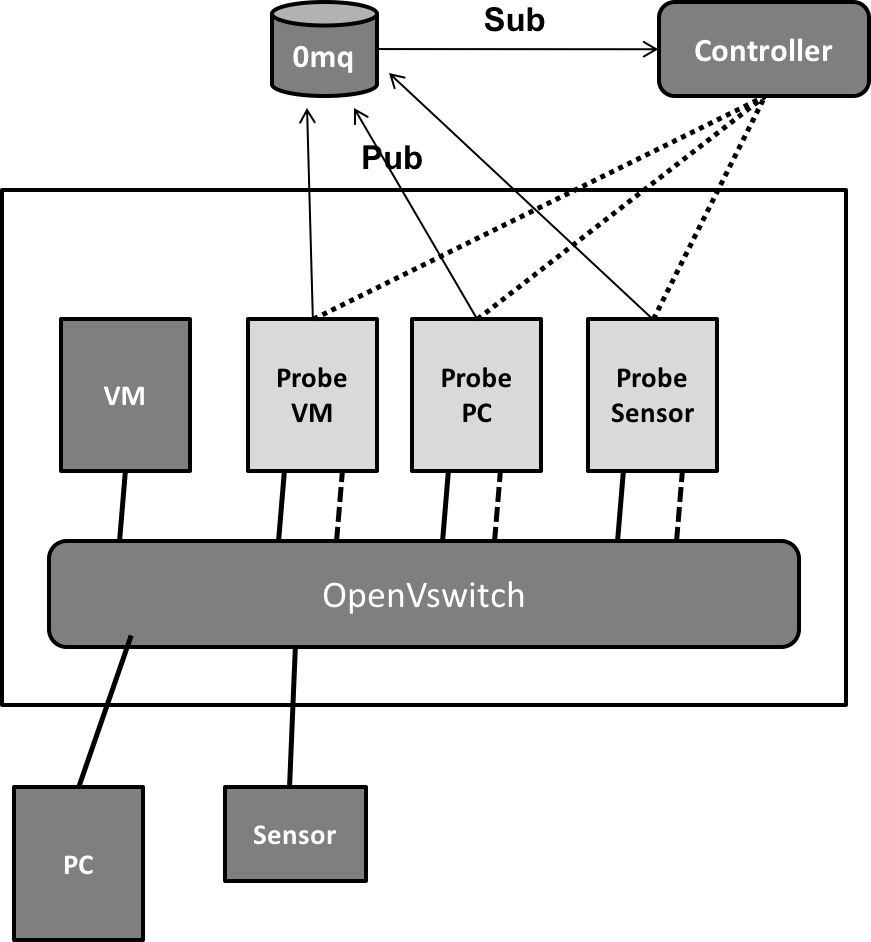}
  \caption{Probes as Containers}
  \label{fig:sub1}
\end{subfigure}
\caption{D-StreaMon architectures}
\vspace{-3ex}
\label{fig:test}
\end{figure}

In this paper, we propose "D-Streamon", a distributed framework for wide-spreading network monitoring based on NFV principles. We call this framework D-Streamon, as for the monitoring engine we used StreaMon~\cite{bianchi2014streamon} an high scalable data-plane programming abstraction for stream-based monitoring tasks directly running inside network probes, which exploits eXtended Finite State Machines (XFSM) for programming custom monitoring logic. The lifecycle of a StreaMon probe has been completely reexamined and StreaMon software has been adapted to run as "containerized" software based probe. 
In this new architecture, the StreaMon probes can communicate between each other and provide information to external frameworks. This can enable advanced monitoring techniques based on the correlation of the information and overcome the problem derived of the single point of control. To complete our framework, we realized a set of management tools which makes agile and flexible the deployment of the StreaMon based probes: the repetitive, not scalable and error prone procedures are replaced by a software based orchestrator. An open source implementation can be downloaded from our repositories~\cite{dstreamonrepo} and is available for evaluation.

\section{D-StreaMon}
\label{sec:dstreamon}

 \begin{figure}
    \centering
    \includegraphics[width=0.48\textwidth]{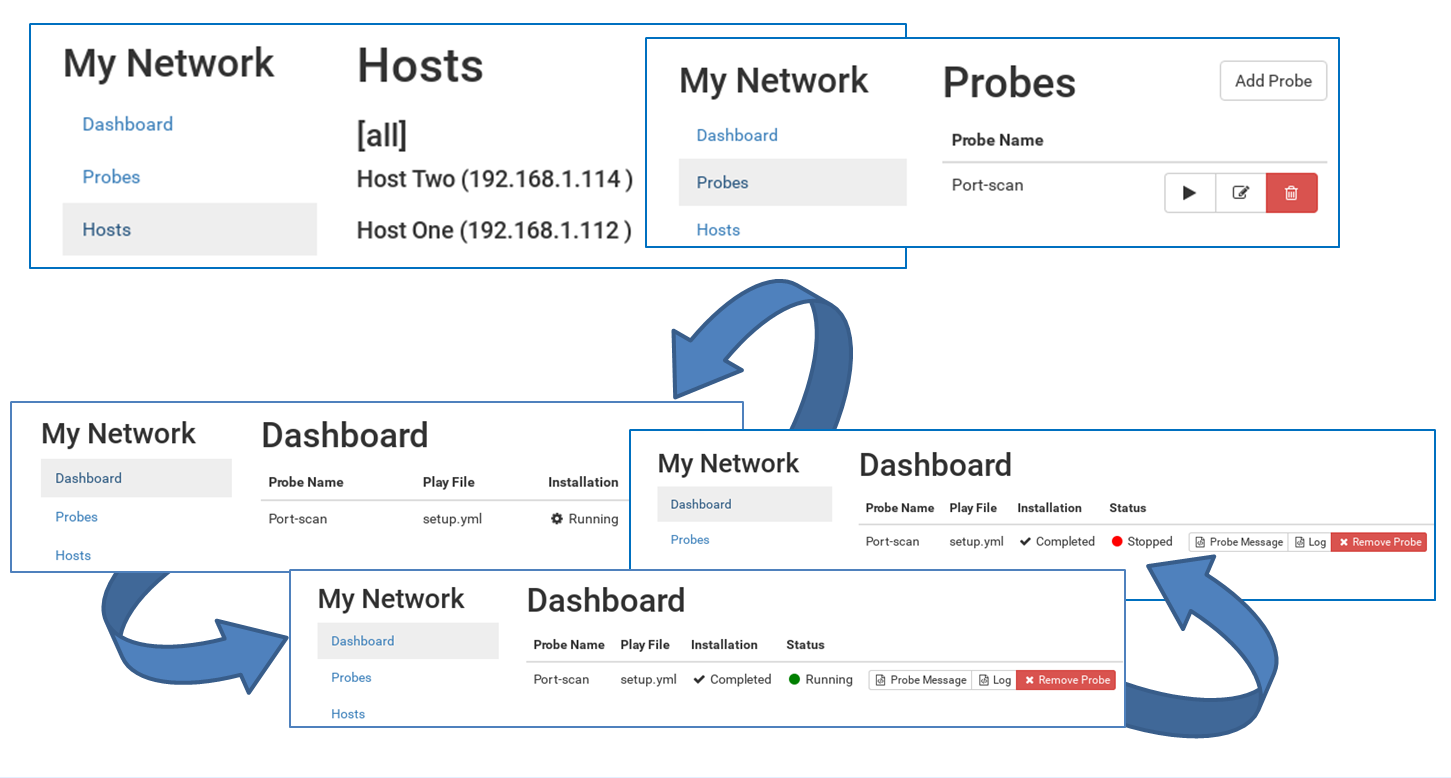}
    \caption{D-StreaMon deployment tools.}
    \vspace{-3ex}
    \label{fig:dstream}
\end{figure}

StreaMon has been originally implemented as specialized middlebox, even though it introduces open abstractions to program the network probe, by design does not consider distributed environment involving an high number of probes. Since it foresees for a vertical integrated stack where \textit{Control} and the \textit{Monitoring Probe} run on the same node, firstly we have introduced a decoupling between them in the new architecture. 
The D-StreaMon's controller is in charge of managing probes' configurations, described by through the StreaMon XML-based language. These high-level configurations are subsequently used to generate the runnable objects to be pushed and executed in the monitoring probes. In this transition, the Control component is committed to perform also the cross-compilation of the StreaMon runnables. This approach can introduces several benefits in the architecture: i) maintain the monitoring process agnostic to the platform that runs the probe; ii) restrict the operational functions to the controller and iii) limit the number of components that runs on the probe. 
The whole management process of the network monitoring probes has been implemented leveraging on the Ansible framework. With no operational overhead and without requiring the presence of agents in the managed hosts, 
it allows to remotely execute tasks in the host. 
To support the operator in the management and the supervision of functionalities described above, D-StreaMon provides an intuitive and simple to use management dashboard (figure \ref{fig:dstream}), 
that enables to perform the installation, execution and removal of monitoring probes in the host machines. As regards the communication architecture between probes and the controller, D-StreaMon does not rely on the classical client/server model. Since the monitoring probes can be hierarchically distributed (to process the network flows at different levels) and should even be able to cooperate, D-StreaMon exploits the flexibility of the publish/subscribe networking model. Each monitoring probe acts as an independent publisher following the actions defined in the configuration of the XFSM and publishes events assigning to them a label. 
It permits to identify the type of information forwarded (e.g. info, alert, warning, log, etc.) and allows interested subscribers to filter events based on that topic. Note that the controller is a \textit{special} subscriber that listens to the whole set of topics created by each monitoring probe. This architecture has been realized by means of the ZeroMQ library. 

Using the architecture described above, we have designed two NFV based deployments. The first solution is referred to as \textit{Probes as Processes} (figure \ref{fig:sub2}). The monitoring Probes are executed as processes running in the main host process space. The Probes intercept the traffic directly from the ports of the target which have been bridged in the L2 switch of the virtualization server (in our case we use Open vSwitch to implement the Layer 2 switch of the virtualization server). The second solution is referred to as \textit{Probes as Containers} and is represented in figure \ref{fig:sub1}. It envisages the use of the Docker Containers for running the probes: for each target  to be monitored, a StreaMon Probe is created and deployed in a Docker Container. All Containers are attached to the same L2 switch and the port mirroring functionality is used to "copy" the VM traffic towards the associated Probe.  The first NFV solution can provide better performance, as the Process virtualization introduces less overhead with respect to Containers solution and because the packets do not need to be "mirrored" by the switch. At the same time the second solution can provide a better isolation and does not require particular workarounds to have different StreaMon Probes working in several Processes co-located in the same host.

Future works will address: i) the improvement of the packet processing performances of the D-StreaMon probes by replacing \textit{Libpcap} with a more performant solution; ii) the realization of a coordinated solution which integrates NFV, SDN and Cloud Computing to have automated deployment of probes and promptly react to malicious actions.





%

\bibliographystyle{IEEEtran}
\bibliography{main}

\end{document}